\documentclass[aps,preprint]{revtex4}

\usepackage{amsmath,amssymb, bbold, mathtools,amsthm}
\usepackage{stmaryrd} 
\usepackage{hyperref}
\usepackage{verbatim} 
\usepackage{tikz}
\tikzset{node distance=3cm, auto}

\def\hil#1{{{#1}}}   
\def\hilm#1{{{#1}}}   
\def\colorred#1{{{#1}}}   

\newtheorem{theorem}{Theorem}[section]
\newtheorem{lemma}[theorem]{Lemma}

\newtheorem{corollary}[theorem]{Corollary}
\newtheorem{proposition}[theorem]{Proposition}

\theoremstyle{definition}

\theoremstyle{remark}

\numberwithin{equation}{section}

\def\bbC{\mathbb{C}}

\def\bbF{\mathbb{F}}

\def\bbZ{\mathbb{Z}}

\def\ba{{\bf a}}
\def\bb{{\bf b}}
\def\bc{{\bf c}}
\def\bd{{\bf d}}

\def\bs{{\bf s}}
 
\def\bv{{\bf v}}
\def\bw{{\bf w}}
\def\bx{{\bf x}}
\def\by{{\bf y}}
\def\bz{{\bf z}}

\def\GFq { \bbF_{q }}

\def\cE{\mathcal{E}}

\def\d{\delta}

\def\w{\omega}

\def\>{\rangle}
\def\<{\langle}
\def\llb{\llbracket}
\def\rrb{\rrbracket}

\DeclareMathOperator{\tr}{Tr}

\DeclareMathOperator{\mindist}{mindist}

\def\Pr{{\rm Pr} }

\newcommand{\bi}[2]{\dbinom{#1}{#2}}

\begin{document}
%
\title{Concatenated quantum codes can attain the quantum Gilbert-Varshamov bound}

\author{Yingkai Ouyang }
\affiliation{
\footnotesize Department of Combinatorics and Optimization, Institute of Quantum Computing, University of Waterloo, \\ 
200 University Avenue West,
\footnotesize Waterloo, Ontario N2L 3G1, Canada.\\ 
\footnotesize \texttt{y3ouyang@math.uwaterloo.ca}}

\begin{abstract}
A family of quantum codes of increasing block length with positive rate is asymptotically good if the ratio of its distance to its block length approaches a positive constant. The asymptotic quantum Gilbert-Varshamov (GV) bound states that there exist $q$-ary quantum codes of sufficiently long block length $N$ having fixed rate $R$ with distance at least $N H^{-1}_{q^2}((1-R)/2)$, where $H_{q^2}$ is the $q^2$-ary entropy function. For $q < 7$, only random quantum codes are known to asymptotically attain the quantum GV bound. However, random codes have little structure. In this paper, we generalize the classical result of Thommesen \cite{Tho83} to the quantum case, thereby demonstrating the existence of concatenated quantum codes that can asymptotically attain the quantum GV bound. The outer codes are quantum generalized Reed-Solomon codes, and the inner codes are random independently chosen stabilizer codes, where the rates of the inner and outer codes lie in a specified feasible region.\end{abstract}

 \maketitle



%

\section{Introduction}

A family of $q$-ary quantum codes \cite{Rai99NQC} of increasing block length with positive rate is defined to be {\it asymptotically good} if the ratio of its distance to its block length approaches a positive constant. 
Designing good quantum codes is highly nontrivial, just as it is in the classical case.  The quantum Gilbert-Varshamov (GV) bound \cite{gottesman-thesis, ABKL00, AKn01, FeM04, MaY08, JiX11} is a lower bound on an achievable relative distance of a quantum code of a fixed rate, and is attainable for various families of random quantum codes \cite{gottesman-thesis, AKn01, MaY08}. Explicit families of quantum codes, both unconcatenated \cite{ALT01, Mat02} and concatenated \cite{CLX01,Fuj06,Ham08,LXW09}, have been studied, but do not attain the quantum GV bound for $q < 7$ \cite{Nie07}. We show that concatenated quantum codes can attain the quantum GV bound. 

We are motivated by the historical development of the idea of concatenating a sequence of increasingly long classical Reed-Solomon (RS) outer codes with various types of \hil{classical} inner codes. In both cases where the inner codes are all identical \cite{sloane} or all distinct \cite{Jus72}, the resultant sequence of concatenated codes while asymptotically good nonetheless fail to attain the GV bound. A special case of Thommesen's result \cite{Tho83} shows that even if the inner codes all have a rate of one, if they are chosen uniformly at random, the resultant sequence of concatenated codes almost surely attains the GV bound. Our work extends this classical observation to the quantum case. 

\begin{figure}[h!]
  \centering
    \includegraphics[width=1\textwidth]{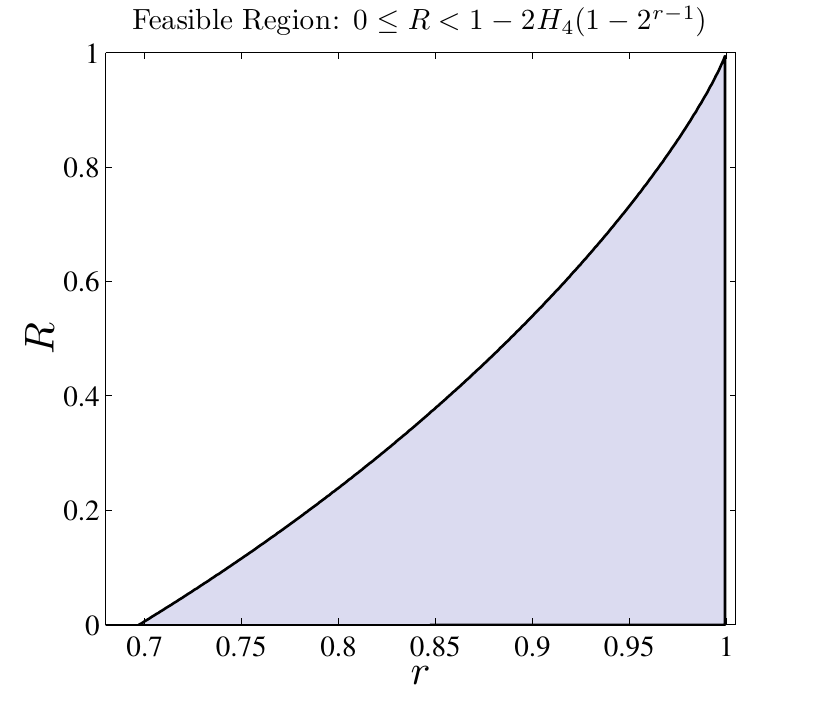}
      \caption{When $p=2$, the shaded region depicts the rates $r$ and $R$ for which Theorem~\ref{thm:qgv} applies.\label{fig:feasible-region}}
\end{figure}

  We show the quantum analog of Thommesen's result -- the sequence of concatenated quantum codes with the outer code being a quantum generalized RS code \cite{GGB99, GBR04,LXW08, LXW09} and random inner stabilizer codes almost surely attains the quantum GV bound when the rates of the inner and outer codes lie in feasible region (\ref{ineq:feasible-condition}) with an example depicted in {Figure~\ref{fig:feasible-region}}. The property of the outer code that we need is that the normalizer of its stabilizer is classical maximal distance separable (MDS) code \cite{LXW08}. Our work is closest in spirit to that of Fujita \cite{Fuj06}, where quantum equivalents of the Zyablov and the Blokh-Zyablov bounds are obtained (not attaining the quantum GV bound) by choosing a quantum RS code with essentially random inner codes.
  
In the proof of the classical result, Thommesen uses a random coding argument to compute the probability that any codeword of weight less than the target minimum distance belongs to the random code. 
Subsequently, he uses the union bound, the spectral property of the Reed-Solomon outer code, and properties of the $q$-ary entropy function \colorred{(defined in \ref{def:qary_entropy})}, to prove that the proposed random code almost surely does not contain any codeword of weight less than the prescribed minimum distance. 

The proof of our quantum result follows a similar strategy, with codewords replaced by elements of the normalizer not in the stabilizer. However the feasible region for the rates of the inner and outer codes for the classical and the quantum result are not analogous, 
because the monotonicity of the $q$-ary entropy function applies in a different feasible region from that of the classical case.

The organization of this paper is as follows: Section \ref{sec:prelims} introduces the notation and preliminary material used in this paper. This section lays out the formalism of concatenating stabilizer codes, which is crucial to the proof of the main result. We state our main result in Theorem~\ref{thm:qgv} of Section \ref{sec:main}, and the \hil{remainder} of the \hil{paper} is dedicated to its proof.

\section{Preliminaries}\label{sec:prelims}
Let $L(\bbC^q)$ denote the set of complex $q$ by $q$ matrices. Define $\mathbb 1_q$ to be a size $q$ identity matrix and $\w_q \coloneqq e^{2\pi i / q}$ to be a primitive $q$-th root of unity, where $q \ge 2$ is an prime power. Define $ 0 \log_q 0 \coloneqq 0$. Define the $q$-ary entropy function and its inverse to be $H_q : [0,1] \to [0,1] $ and $H_q^{-1}:[0,1] \to [0,\frac{q-1}{q}]$ respectively where
\begin{align}
H_q(x) &\coloneqq x \log_q (q-1) - x \log_q x - (1-x) \log_q (1-x). 
\label{def:qary_entropy}
\end{align}
\colorred{
The $q$-ary entropy function is important here because it helps us to count the size of sets with $q$ symbols. The base-$q$ logarithm of the number of vectors from $\bbF_q^n$ that differ in at most $xn$ components from the zero-vector is dominated by $n H_q(x)$ as $n$ becomes large. 
}

\colorred{
For a ground set $\Omega$ and $n$-tuples ${\bf x} \in \Omega^n$, define $x_j$ to be $j$-th element of the $n$-tuple ${\bf x}$. Given tuples ${\bf x} \in \Omega^n$ and ${\bf y} \in \Omega^m$, define the pasting of the tuples ${\bf x}$ and ${\bf y}$ to be $({\bf x} | {\bf y})\coloneqq (x_1 ,\ldots, x_n , y_1, \ldots , y_m )$. When $M_1$ and $M_2$ are matrices with the same number of columns, define 
$
(M_1; M_2) \coloneqq 
\begin{pmatrix}
M_1 \\ M_2 \\
\end{pmatrix}.
$
For positive integer $\ell$, define $[\ell] \coloneqq \{1,\ldots ,\ell\}$.
Define the Hamming distance $d_H(\bx , \by)$ between ${\bf x} \in \Omega^n$ and $\by \in \Omega^n$ as the number of indices on which $\bx$ and $\by$ differ. Define the minimum distance of any subset $C \subset \Omega^n$  to be
$
\mindist(C)\coloneqq \min_{\bx,\by \in C} \{d_H(\bx, \by) : \bx \neq \by \}.
$
}

\colorred{A code over a vector field $\bbF_q^n$ is $q$-ary linear code of length $n$ if it is a subspace of $\bbF_q^n$.}
An additive code is a subgroup of the field under the field addition operation.
A classical $q$-ary linear code \cite{sloane} of block length $n$ and $k$ generators with minimum distance of $d$ is said to be an $[n,k]_q$ code or an $[n,k,d]_q$ code. 
A classical $[n,k,d]_q$ code is maximally distance separated (MDS) if $d = n-k+1$.
A quantum $q$-ary stabilizer code \cite{Rai99NQC} of block length $n$ encoding $k$ qudits is said to be an $\llb n,k \rrb_q$ code. The rates of an $\llb n,k\rrb_q$ code and an $[n,k]_q$ code are both defined to be $\frac{k}{n}$.
\subsection{Finite Fields and $q$-ary Error Bases} \label{subsec:qaryPaulis}
We briefly review $q$-ary error bases \cite{AKn01}.
Given a prime number $p$, let $q = p^k$ where $k$ is a positive integer. 
Let generalizations of the qubit Pauli matrices be
\begin{align}
X &\coloneqq   \sum_{j=0}^{p-1} |(j + 1)\!\! \mod p \>\<j| \notag\\
Z &\coloneqq   \sum_{j=0}^{p-1} (\w_p)^j |j\>\<j|
\end{align}
which satisfy the commutation property 
$X^a Z ^b= (\w_p)^{ab} Z^b X^a $
for non-negative integers $a$ and $b$. 
 We define the matrix 
\begin{align}
X_{\ba} Z_{\bb} 
\coloneqq X^{a_1} Z^{b_1} \otimes \ldots \otimes  X^{a_k} Z^{b_k}
\end{align}
 as a single qudit $q$-ary error basis element. 
We define a $q$-ary error basis on a single qudit as the set 
 $
   \cE_q \coloneqq \{ X_\ba Z_\bb : \ba,\bb \in\bbZ_p^k  \}.
 $
A $q$-ary error basis on $n$ qudits is defined as 
$\cE_q ^{\otimes n}$
and its basis elements have the form
\[
X_{\ba^{(1)}} Z_{\bb^{(1)}}  \otimes \ldots \otimes X_{\ba{(n)}} Z_{\bb^{(n)}} = 
 X_{(\ba^{(1)}|\ldots | \ba^{(n)})} Z_{(\bb^{(1)}| \ldots |\bb^{(n)})} .
 \]
Now let $t$ be any positive integer. \
Observe that for 
 $\ba, 
  \bb ,
    \bc ,
      \bd \in \bbZ_p^t$,
  the matrices $X_\ba Z_\bb$ and $X_{\bc} Z_{\bd}$ satisfy the commutation relation
 \[
( X_{\ba} Z_\bb)( X_{\bc} Z_{\bd})
=( X_{\bc} Z_{\bd})( X_{\ba} Z_\bb) 
(\omega_p)^{\sum_{i=1}^t a_i d_i - b_i c_i }.
 \]
Hence the symplectic scalar product 
\[
\< (\ba |\bb) , (\bc|\bd) \>_s \coloneqq \sum_{i=1}^t a_i d_i - b_i c_i = \ba \bd^T - \bb \bc^T 
\]
quantifies the commutation relation between the matrices $X_\ba Z_\bb$ and $X_{\bc} Z_{\bd}$. 
When this scalar product is zero, we say that the vectors $(\ba|\bb)$ and $(\bc|\bd)$ are $s$-orthogonal, and the matrices
$X_\ba Z_\bb$ and $X_\bc Z_\bd$ 
commute under matrix mutiplication.

We now elucidate the connection between $q$-ary error bases and finite fields.
Define the trace function from the field $\bbF_q$ to $\bbF_p$ to be $\tr: x \mapsto \sum_{i=0}^{k-1} x^{p^i}$.
Also let $\{ \gamma, \gamma^q\}$ be a basis of $\bbF_{q^2}$ over $\bbF_q$,
where $\gamma$ and $\gamma^q$ are the distinct roots of an irreducible degree-2 polynomial over $\bbF_q$.
Now let $\mathbb a \coloneqq \hilm{(} \alpha_1, \ldots, \alpha_k )$
and $\mathbb b \coloneqq ( \beta_1, \ldots, \beta_k )$ be dual bases of $\bbF_q$ so that 
${\mathbb a}^T{\mathbb b}$ is a size $k$ identity matrix. 
Also let $\ba, \bb,\bc$, and $\bd$ be vectors from 
$\bbZ_p^k$. 
Then $
\tr((\ba \mathbb a^T)(\bb \mathbb b^T)) 
= 
\tr(\ba \mathbb a^T \mathbb b \bb^T)) 
  = \ba \bb^T,
$
which implies that 
\begin{align}
\tr(
(\ba \mathbb a^T)(\bd \mathbb b^T)
	-
(\bb \mathbb a^T)(\bc \mathbb b^T)
) 
  = \ba \bd^T - \bb \bc^T.\label{eq:tr-sympletic}
\end{align}
Given the vectors $\bx$ and $\by$ in $\bbF_{q^2}^n$, the Hermitian scalar product (see (28) of \cite{AKn01}) between $\bx$ and $\by$ is 
\[
\< \bx, \by \>_h \coloneqq \sum_{i=1}^n (x_i)^q y_i.
\] 
When this Hermitian scalar product is zero, we say that $\bx$ and $\by$ are $h$-orthogonal.
This scalar product is called Hermitian because taking an element of $\bbF_{q^2}$ to the $q$-th power is analogous to conjugation over the complex field. 
For any subset $C \subset \bbF_{q^2}^n$, we also define its Hermitian dual to be 
$C^{\perp_h} \coloneqq \{ \by \in \bbF_{q^2}^n: \<\bx,\by\>_h = 0 , \bx \in C\}$. 

\begin{figure}
\begin{center}
\begin{tikzpicture}
  \node (C) {$W \in \cE_q^{\otimes n}$};
  \node (P) [below of=C] {$\bw \in \bbF_q^{2n}$};
  \node (Ai) [right of=P] {$\tilde \bw \in \bbF_{q^2}^n$};
  \draw[->] (C) to node {{\Large $\tilde \varphi$}} (Ai);
  \draw[->] (C) to node [swap] {{\Large $\varphi$}} (P);
  \draw[->] (P) to node [swap] {{\Large $\psi$}} (Ai);
\end{tikzpicture}
\caption{Equivalent representations of an $n$-qudit $q$-ary error basis element. \label{fig:qary-representations}}
\end{center}
\end{figure}
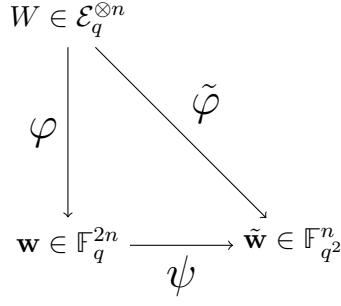

The following proposition 
shows that if two error basis elements are to commute, it suffices for their $q^2$-ary finite field counterparts to be $h$-orthogonal.
\begin{proposition}[\cite{AKn01}]\label{prop:hermitian-dual}
Let $\bx,\by \in \bbF_{q^2}^n$, and suppose that $\<\bx, \by\>_h = 0$.
For all $i \in [n]$, let $x_i$ and $y_i$ have the decompositions
\begin{align*}
x_i = x_{i,1}\gamma+ x_{i,2}\gamma^q 
	&= \ba^{(i)}\mathbb a^T \gamma + 
	   \bb^{(i)}\mathbb b^T  \gamma^q,\\
y_i = y_{i,1}\gamma+ y_{i,2}\gamma^q 
	&= \bc^{(i)}\mathbb a^T \gamma+ 
	   \bd^{(i)}\mathbb b^T  \gamma^q,
\end{align*}
where 
$x_{i,1}, x_{i,2}, y_{i,1}, y_{i,2} \in \bbF_q$ 
 and $\ba^{(i)},\bb^{(i)},\bc^{(i)}, \bd^{(i)} \in \bbZ_p^k$.
 Then the matrices 
$
X_{(\ba^{(1)} | \ldots | \ba^{(n)})}
Z_{(\bb^{(1)} | \ldots | \bb^{(n)})}
$ 
and 
$
X_{(\bc^{(1)} | \ldots | \bc^{(n)})}
Z_{(\bd^{(1)} | \ldots | \bd^{(n)})}
$ 
from the set $\cE_q^{\otimes n}$ commute under matrix multiplication.
\end{proposition}
\begin{proof}
Since $\<\bx, \by\>_h = (\<\by, \bx\>_h)^q$ and $0^q = 0$, we have 
$\<\bx, \by\>_h= 0$ implying that $\<\by, \bx\>_h = 0$. Thus 
$\<\bx, \by\>_h - \<\by, \bx\>_h = 0$, which implies that
$\sum_{i=1}^n x_i^q y_i - y_i^q x_i=0 $.
Hence
\begin{align}
0 &= \sum_{i=1}^n 
    \left( 
    (x_{i,1}\gamma^q + x_{i,2}\gamma)  (y_{i,1}\gamma + y_{i,2}\gamma^q)  
    -
    (y_{i,1}\gamma^q + y_{i,2}\gamma)  (x_{i,1} \gamma + x_{i,2}\gamma^q)  
    \right) 
    \notag\\
&= (\gamma-\gamma^2)\sum_{i=1}^n 
    (x_{i,1}y_{i,2} - x_{i,2}y_{i,1}).
\end{align}
If $\gamma = \gamma^2$, then $\gamma = \gamma^q$ which is a contradiction. Hence $\gamma \neq \gamma^2$ which implies that 
\[
\sum_{i=1}^n 
    (x_{i,1}y_{i,2} - x_{i,2}y_{i,1}) = 0.
\]
Let ${\ba = (\ba^{(1)} | \ldots | \ba^{(n)}) }$, 
${\bb = (\bb^{(1)} | \ldots | \bb^{(n)})}$, 
${\bc = (\bc^{(1)} | \ldots | \bc^{(n)})}$, and
${\bd = (\bd^{(1)} | \ldots | \bd^{(n)})}$. 
Tracing both sides of the above equation gives ${\< (\ba |\bb), (\bc | \bd) \>_s = 0}$, which implies that the matrices 
$
X_{\ba} Z_{\bb}
$ 
and 
$
X_{\bc} Z_{\bd}
$ commute.
\end{proof}
In view of Proposition~\ref{prop:hermitian-dual} and (\ref{eq:tr-sympletic}), each element of a $q$-ary error basis over $n$ qubits 
$W= X_{(\ba^{(1)}|\ldots|\ba^{(n)})} 
 Z_{(\bb^{(1)}|\ldots|\bb^{(n)})} $
 can be represented by the codewords 
 $\varphi(W) \coloneqq\bw \in \bbF_q^{2n}$ 
 and
  $\tilde \varphi(W) \coloneqq\tilde \bw \in \bbF_{q^2}^n$, 
  where for $i \in [n]$, 
 \begin{align*}
 w_i     &= \ba^{(i)} \mathbb a^T,\notag\\
 w_{i+n} &= \bb^{(i)} \mathbb b^T,\\
 \tilde w_i &= w_i \gamma + w_{i+n} \gamma^q .
 \end{align*}
We define the map $\psi$ to take $\bw$ to $\tilde \bw$. 
Let the maps $\psi, \varphi$ and $\tilde \varphi$ act component-wise on sets and matrices. Consequently, elements of an error basis can be studied in their different finite field representations, with the bijective maps $\varphi, \tilde \varphi$ and $\psi$ depicted in Figure~\ref{fig:qary-representations}.
\subsection{Stabilizer Codes}\label{subsec:stabilizercodes}
Given a prime number $p$, let $q = p^k$ where $k$ is a positive integer. 
Given a subset $S \subset \cE_q^{\otimes n}$ where $\varphi(S)$ is an additive group with $s$ independent additive generators, the maximal subspace of  $(\bbC^q)^{\otimes n}$ left invariant under the action of all elements of $S$ is called an 
$\llb n,n-\frac{s}{k} \rrb_q$
 stabilizer code. 
 The sets $S$, $\varphi(S)$ and $\tilde \varphi(S)$ 
 are the stabilizers of our stabilizer code in the matrix representation, the $\bbF_q^{2n}$-representation and the $\bbF_{q^2}^n$-representation respectively.
We study stabilizer codes in the language of 
finite fields \cite{gottesman-thesis, AKn01}.  

Consider the full rank generator matrix 
$G = (G_{\rm stb};  G_{\rm x} ; G_{\rm z})$
over $\GFq$ with $(2kn-s)$ rows and $2n$ columns where
the stabilizer generator 
${G_{\rm stb} = (\bs^{(1)}; \ldots ; \bs^{(s)})}$, the logical-X generator
 ${G_{\rm x}  = (\bx^{(1)}; \ldots ; \bx^{(kn-s)})}$, and the logical-Z generator 
  ${G_{\rm z} = (\bz^{(1)}; \ldots ; \bz^{(kn-s)})}$ are submatrices of G.
We also require $G = (G_{\rm stb};  G_{\rm x} ; G_{\rm z})$ to have the properties:
\begin{enumerate}
\item Each row of $G_{\rm stb}$ is $s$-orthogonal to every row of $G$.
\item For all $i,j \in [kn-s]$, 
$\< \bx^{(i)} , \bz^{(i)} \>_s 
= \delta_{i,j}$, where $\delta_{i,j}$ is the Kronecker delta.
\end{enumerate}
The error basis elements corresponding to the rows of $G_{\rm x}$ and $G_{\rm z}$ are generators for logical operations that can be applied on the stabilizer code.

We denote the additive (not necessarily linear) classical codes generated by $G_{\rm stb}$ and $G$ under field addition by $C_{\rm stb}$ and $C_{\rm nrm}$ respectively.
The set of all elements in $\bbF_{q}^{2n}$ that are $s$-orthogonal to all elements in $C_{\rm stb}$ is $C_{\rm nrm}$. 
The minimum distance of our stabilizer code is the minimum distance 
of the punctured code 
$\tilde C_{\rm pnc} \coloneqq \{ x \in \psi( C_{\rm nrm}): x \notin \psi(C_{\rm stb}) \}$ \cite{AKn01}. 
 We denote an 
 $\llb n,n-\frac{s}{k} \rrb_q$ 
 stabilizer code with distance $d$ as 
 $\llb n,n-\frac{s}{k}, d \rrb_q$. 
 The rate of the stabilizer code is $1 - \frac{s}{k n}$ and its relative distance is $\frac{d}{n}$.
 
We define a {\it random $\llb n, n-\frac{s}{k} \rrb_q$ stabilizer code} to be a stabilizer code corresponding to a generator matrix $G = (G_{\rm stb}; G_{\rm x}; G_{\rm z})$ chosen uniformly at random from all possible generator matrices with 
$(2kn - s)$ rows and $2n$ columns over the vector field $\bbF_q^{2n}$.

Let the rates and relative distances of an infinite code sequence of $\{ \llb n,n r_n,n \d_n \rrb_q \}_{n }$ converge to the positive numbers $r$ and $\delta$ respectively.   
If 
\begin{align}
\delta \ge H_{q^2}^{-1}\left(\frac{1-r}{2}\right) \label{def:quantum-gv},
\end{align}
we say that the code sequence
attains the asymptotic quantum $q$-ary GV bound.


\subsection{Concatenation of Stabilizer Codes} \label{subsec:concat}

Concatenation makes a longer code from an appropriately chosen set of shorter codes.
We consider only the concatenation of stabilizer codes.
Let $q = p^k$ where $p$ is prime.

The quantum message that we wish to encode into a concatenated quantum code is a $q^K$-dimension 
 quantum state which is first encoded into an $\llb N,K \rrb_{q}$ {\it outer code}. Let our $\llb N,K\rrb_q$ outer code be generated by
${G^{\rm (out)} = 
(G^{\rm (out)}_{\rm stb};
 G^{\rm (out)}_{\rm x};
 G^{\rm (out)}_{\rm z} )}$. The outer code comprises of $N$ blocks of dimension $q$ complex Euclidean spaces, with each of these $N$ blocks further encoded as an $\llb n,k \rrb_{p}$ {\it inner code}. 
 Let the $j$-th $\llb n,k \rrb_{p}$ inner code be generated by 
 $G^{ (j)} = 
 (G_{\rm stb}^{ (j)};
  G_{\rm x}^{ (j)}; 
  G_{\rm z}^{ (j)} )$, with 
 $G_{\rm x}^{ (j)} = (\bx^{(j),1}; \ldots ; \bx^{(j),k}) 
  $ 
  and
 $G_{\rm z}^{ (j)}  = (\bz^{(j),1}; \ldots ; \bz^{(j),k})$
 for $j \in [N]$. 
 The resultant code is a concatenated code with parameters $\llb nN, kK \rrb_{p}$， 
 generated by 
$G^{\rm (concat)} = 
(G_{\rm stb}^{\rm (concat)};
 G_{\rm x}^{\rm (concat)}; 
 G_{\rm z}^{\rm (concat)} )$. 
 
We now elucidate the construction of the generator of the concatenated code $G^{\rm (concat)}$ using the generator of the outer code
$G^{\rm (out)}$ and the generators of the inner codes $G^{ ( j)}$ for $j \in [N]$.

Using the notation defined in Section \ref{subsec:qaryPaulis}, let the letter $w \in \bbF_{q^2}$ have the decomposition $w = \ba \mathbb a^T \gamma + \bb \mathbb b^T \gamma^q$ where $\ba, \bb \in \bbZ_p^k$.
We define the image of $w$ over the smaller field $\bbF_{p^2}$ with respect to the $j$-th inner code to be the $\psi( C_{\rm stb}^{(j)})$-coset representative given by 
\begin{align}
\pi^{(j)}(w)
\coloneqq&
\sum_{\ell=1}^k \left( 
 a_\ell \bx^{(j),\ell} 
 +  b_\ell \bz^{(j),\ell}
\right).
\end{align}
Given vectors
$\bs  \in [N]^m$ and $\bw \in \bbF_{q^2}^m$, we define 
$
\pi^{\bs}( \bw )
\coloneqq (\pi^{(s_1)}(w_1)
 | \ldots |
  \pi^{(s_m)}(w_m) ) .
$
As a shorthand we define $\pi \coloneqq \pi^{(1, \ldots ,N)}$. 
Let $\pi$ also act component-wise on both matrices and sets.
Then the $\bbF_{p^2}$-representations of the stabilizer generator, the X-logical generator and the Z-logical generator of our concatenated code are given by
\begin{align}
\psi(  G_{\rm stb}^{({\rm concat})})
& = 
\left( \pi (\psi( G_{\rm stb}^{({\rm out})}));
 \begin{pmatrix}
 \psi( G_{\rm stb}^{(1)}) & {\bf 0 }& {\bf 0} &  {\bf 0} \\ 
 {\bf 0 }&   \psi( G_{\rm stb}^{(2)}) &  {\bf 0} & {\bf 0} \\
  {\bf 0 }& {\bf 0} &  \ddots & {\bf 0} \\
   {\bf 0 }& {\bf 0} &  {\bf 0} &  \psi( G_{\rm stb}^{(N)}) \\
 \end{pmatrix}
 \right)
 \notag\\
 \psi (G_{\rm x}^{({\rm concat})})
 &= 
  \pi( \psi( G_{\rm x}^{({\rm out})})), \quad 
 \psi( G_{\rm z}^{({\rm concat})})
 = 
  \pi( \psi( G_{\rm z}^{({\rm out} )})) 
\end{align}
respectively.
The $\bbF_{p^2}$-representations of the stabilizer and the normalizer of the concatenated code are
$
\psi(   C^{\rm (concat)}_{\rm stb} ) \coloneqq 
     \pi(\psi( C_{\rm stb}^{\rm (out)})) + \psi(  C_{\rm stb}^{(1)} \times \ldots. \times 
  C_{\rm stb}^{(N)})
     $
     and
     $
{\psi( C^{\rm (concat)}_{\rm nrm} ) \coloneqq 
     \pi(\psi( C_{\rm nrm}^{\rm (out)}) ) + \psi (C_{\rm stb}^{\rm (concat)})}
$ respectively. 

In this paper, we use some of the $q$-ary quantum codes of Li, Xing and Wang \cite{LXW08} as the outer codes of our concatenated codes. 
The stabilizers and normalizers of these codes are classical MDS codes in the $\bbF_{q^2}$-representation, which is not necessarily the case for other quantum codes \cite{GBR04}.
\begin{theorem}[Li, Xing, Wang \cite{LXW08} \label{thm:lxw}]
Let $N$ be a prime power and $K$ be an even integer in $[0,N]$ such that $\frac{N-K}{2}$ is also an integer. Then
there exists a quantum generalized Reed-Solomon code with parameters 
$\llb N, K, \frac{N-K}{2} +1 \rrb_N$. 
Moreover, the stabilizer $\psi( C_{\rm stb})$ and normalizer $\psi( C_{\rm nrm})$ of this code in the $\bbF_{N^2}$-representation are classical generalized Reed-Solomon codes (are hence classical MDS codes), with $\psi ( C_{\rm nrm}) = \psi (C_{\rm stb})^{\perp_h}$.
\end{theorem}

\section{The Main Result}\label{sec:main}
Our main result is that our sequence of concatenated $p$-ary quantum codes asymptotically attains the quantum GV bound. The outer code is a quantum generalized RS code with $\psi ( C_{\rm nrm}) = \psi (C_{\rm stb})^{\perp_h}$ given by \cite{LXW08}, and the inner codes are independently chosen random stabilizer codes. Theorem~\ref{thm:qgv} is our main result.
\begin{theorem}\label{thm:qgv}
Let $r,R \in \mathbb Q \cap [0,1]$ be the rates of the inner and outer code respectively. Let $p$ be a prime number and $n$ be a positive integer such that $rn$, $N=p^{rn}$, and $\frac{1-R}{2}N \in \bbZ$ are also integers. Also suppose that 
\begin{align}
R < \min \left \{ 1-2H_{p^2} (  1-p^{r-1} ) , 1\right\}. \label{ineq:feasible-condition}
\end{align}
Let $\llb nN, r R n N  , d \rrb_p$ be a concatenated quantum code with a $\llb N, R N  \rrb_{N} $ outer code of given by Theorem~\ref{thm:lxw} concatenated with $N$ independent and identically distributed random $\llb n, r n \rrb_p$ inner quantum codes. Then with probability at least 
 $1- 
 \frac{1}{p^2-1} 
p^{-2N (\frac{1-R}{2})} $,
\[
\frac{d}{nN} > H_{p^2}^{-1} \left( \frac{1-r R}{2} \right) -  \frac{3c(p^2, \frac{1+r}{2}) }{2n}
\]
where $c(p^2,\frac{1+r}{2})$ is a continuity constant as defined in the Appendix in equation (\ref{eq:continuity-constant}).
\end{theorem}

\begin{corollary}
Let $p$ be a prime and $r,R \in [0,1]$ such that the inequality (\ref{ineq:feasible-condition}) holds. For all positive integers $n$, let 
${k_n = \lceil n r \rceil}$, 
${N_n = p^{k_n} }$ and
${K_n = N_n - 2 \lceil \frac{1-R}{2} \rceil}$.
Let $C_n$ be a code formed by concatenating an 
${\llb N_n, K_n \rrb_{N_n}}$ outer code given by Theorem~\ref{thm:lxw} with $N_n$ independent and identically distributed random 
${\llb n, k_n \rrb_p}$ stabilizer codes. Then the code sequence $\{ C_n\}_{n \in \bbZ^+}$ asymptotically attains the quantum Gilbert-Varshamov bound.
\end{corollary}

We proceed to introduce Proposition~\ref{prop:prob-coset} and Lemma~\ref{lem:pr-y-in-utilde}, which are used in the random coding aspects of the proof of Theorem~\ref{thm:qgv}.
\begin{proposition}\label{prop:prob-coset}
Let $\bw$ be any nonzero element of $ \bbF_{p^2}^n$.
Let 
$\psi(C_{\rm nrm})$ and $\psi (C_{\rm stb})$ be the
normalizer and stabilizer over $\bbF_{p^2}$ of 
a random $\llb n,k \rrb_p$ stabilizer code,
and let the corresponding punctured code be 
${\tilde C_{\rm pnc} \coloneqq \{\bw \in \psi( C_{\rm nrm}) : \bw \notin \psi(C_{\rm stb} )\}
}$.
Then ${\Pr[\bw \in \tilde C_{\rm pnc}] < p^{-(n+k)}}$.
\end{proposition}
\begin{proof}
Let $U \subset \bbF_{p}^{2n}$ be a set of independent mutually $s$-orthogonal vectors. Then the number of vectors in $\bbF_p^{2n}$ that are $s$-orthogonal to all elements of $U$ is $p^{2n-|U|}$. Hence 
${\Pr[\bw \in \psi( C_{\rm nrm})] = 
\frac
{\prod_{i=0}^{n-k-1} (p^{2n-1-i} - p^i)}
{\prod_{i=0}^{n-k-1} (p^{2n-i} - p^i)}
< p^{-(n-k)}.}
$ 
The number of cosets of $ C_{\rm stb}$ in $ C_{\rm nrm}$ distinct from $ C_{\rm stb}$ is $p^{2k}-1$. Hence $\Pr[\bw \in \tilde C_{\rm pnc}] < p^{-(n-k)} \frac{1}{p^{2k}-1} < p^{n+k}$.
\end{proof}
\begin{lemma}\label{lem:pr-y-in-utilde}
Let ${\bf W}$ be any nonzero vector in ${\bbF_{q^2}^N}$ of weight $w$, and $h$ be a positive integer no greater than $\frac{p^2-1}{p^2} nw$. Let 
$S= \pi({\bf W})+\psi( C_{\rm stb}^{(1)} \times \ldots \times C_{\rm stb}^{(N)})$ be a random coset.
Then 
${\Pr\left[{\rm minwt }(S) \le h \right] <
(p^2)^{nw H_{p^2} (\frac{h}{nw})}p^{-(n+k)w}.}$
\end{lemma}
\begin{proof}
The minimum weight of 
$S$ is equal to the minimum weight of the random coset
$S' = \pi((W_1,...,W_w))
+
\psi(  C_{\rm stb}^{\rm(1)}\times \ldots \times C_{\rm stb}^{\rm(w)})$, 
where $W_1, \ldots, W_w$ are the nonzero letters of ${\bf W}$.
When $h \le \frac{p^2-1}{p^2}nw$, there are at most $(p^2)^{nw H_{p^2} (\frac{h}{nw})}$ members of $\bbF_{p^2}^n$ of weight no more than $h$ (see \cite{sloane}). 
Let $\bw =(\bv_1| \ldots | \bv_w)$ be any such member of $\bbF_{p^2}^{nw}$, where $\bv_1, \ldots , \bv_w \in \bbF_{p^2}^n$. 
If $\bw$ is also an element of $S'$, each $\bv_i$ is necessarily an element of the non-trivial random coset 
$ \pi^{(i)}(W_i)+\psi(C_{\rm stb}^{(i)})$, the probability of which is less than $p^{-(n+k)}$ by the Proposition~\ref{prop:prob-coset}. Hence the probability that 
$\bw$ is an element of the random set $S'$ is less than $p^{-(n+k)w}$. 
Subsequently, applying the union bound on the number of $\bw$ with a weight no more than $h$ gives the result.
\end{proof}
Now we proceed to prove our main result, Theorem~\ref{thm:qgv}.
\begin{proof}[Proof of Theorem~\ref{thm:qgv}]
To prove our main result, we have to find a designed distance $h > 0$ such that:
\begin{enumerate}
\item The probability that the distance of our concatenated quantum code is less that $h$ is negligible.
\item The designed relative distance $\frac{h}{nN}$ asymptotically attains the quantum GV bound.
\end{enumerate}
We first determine a sufficient condition for $\Pr[d \le h]$ to vanish as $n$ becomes large. 
Now our outer code's normalizer $\psi(C_{\rm nrm}^{\rm(out)})$ is a classical MDS code \cite{LXW08} with parameters 
$[N,N R_{\rm nrm},D]_{q^2}$ where
$D = N(1 -R_{\rm nrm})+1$ and $R_{\rm nrm} \coloneqq \frac{1+R}{2}$. 
The MDS property of our outer code's normalizer implies that the spectrum of the normalizer $A_w$, defined as the number of codewords in 
$\psi( C_{\rm nrm}^{\rm(out)})$ with weight $w \in [D,N]$, is at most 
$
\bi{N}{w} (p^{2k})^{w-D+1}\label{ineq:spectrum}
$
 (see the references \cite{sloane,Tho83}).
 Let 
 ${ \tilde  C_{\rm pnc}^{\rm(concat)}
     \coloneqq \{ {\bf W} \in \psi( C_{\rm nrm}^{(\rm concat)}): {\bf W} \notin \psi( C_{\rm stb}^{(\rm concat)} )\}}$.
Our upper bound on the spectrum $A_w$, the union bound and Lemma~\ref{lem:pr-y-in-utilde} imply that
\begin{align*}
\Pr[d \le h] &= 
\Pr[
{\rm minwt}( \tilde  C_{\rm pnc}^{\rm(concat)}) \le h] \notag\\
&\le
    \sum_{
    \substack{
    {\bf W} \in \psi( C_{\rm nrm}^{\rm(out)}) \\
    {\bf W} \neq 0 \\
    }}
    \Pr\left[
    {\rm minwt }( \pi( {\bf W})
    + \psi ( C_{\rm stb}^{(1)}) \times \ldots \times C_{\rm stb}^{(N)})) \le h
    \right]      \\
 &<
     \sum_{w=D}^{N} 2^N (p^{2k})^{w-D'+1}
(p^2)^{nw H_{p^2} (\frac{h}{nw}) - \frac{n+k}{2} w} 
\le \sum_{w=D}^{\infty} (p^2)^{-nw \eta},
\end{align*}
where
\begin{align}
\eta &\coloneqq - \frac{N}{2nw} -  r \left(1 -\frac{D}{w} + \frac{1}{w}\right) - H_{p^2} \left(\frac{h}{nw}\right) + \frac{1+r}{2}  \label{eq:gamma}.
\end{align}
Now let $\theta = 1 -\frac{D}{w} + \frac{1}{w}$ and observe that $0 \le \theta < R_{\rm nrm}$ for our feasible values of $w$. 
If 
$\eta \ge \frac{1}{n}$ for all $w \in [D, N]$, then 
$\Pr[d \le h] \le
     (p^{2})^{-D}\frac{1}{1-p^{-2 }}.
$
We will determine feasible values of the designed distance $h$ for which the inequality
$\eta \ge \frac{1}{n}$ holds.

Since the inverse entropy function is monotone increasing on the open unit interval, it suffices to require that our choice of $h$ satisfies the inequality
\begin{align}
  \frac{h}{nN}
  &\le \frac{w}{N} H_{p^2}^{-1}  \left( \frac{1+r}{2} -  r \theta - \frac{N}{2nw}  - \frac{1}{n} \right) \label{eq:h-in-terms-of-gamma}.
\end{align}
It suffices to have $\frac{h}{nN}$ equal to some lower bound on the right hand side of the inequality (\ref{eq:h-in-terms-of-gamma}).
Continuity of the inverse entropy (Lemma~\ref{lem:nonincreasing}) and the substitution $\frac{w}{N} = \frac{1-R_{\rm nrm}}{1-\theta} $ gives
\begin{align}
         & \frac{1-R_{\rm nrm}}{1-\theta} H_{p^2}^{-1}  \left( \frac{1+r}{2} -  r \theta - \frac{1}{n} \left(\frac{N}{2w}+1 \right) \right) \notag\\
  \ge    &
         \frac{1-R_{\rm nrm}}{1- \theta} H_{p^2}^{-1}  \left( \frac{1+r}{2} - r \theta   \right) - \left(\frac{1}{2} + \frac{w}{N} \right)  \frac{c(p^2,\frac{1+r}{2} - r \theta)}{n} \label{ineq:final}.
\end{align}
The inequality (\ref{ineq:feasible-condition}) together with our restriction that ${r,R\in[0,1]}$ imply that $r$ and $R$ satisfy the requirements of Lemma~\ref{lem:nonincreasing}. Hence Lemma~\ref{lem:nonincreasing} implies that 
${\frac{1-R_{\rm nrm}}{1-\theta} H^{-1}_{p^2} \left( \frac{1+r}{2} - r \theta \right)}$ is a monotonic non-increasing function of $\theta$.
Since ${\frac{c(p^2,\frac{1+r}{2} - r \theta)}{n}}$ is also a monotonic non-increasing function of $\theta$ for feasible values of $r$ and $R$, the right hand side of (\ref{ineq:final}) 
is at least
$ H^{-1}_{p^2} \left( \frac{1 - r R}{2} \right) - 
\frac{3c(p^2, \frac{1+r}{2}) }{2n}  $
by setting $\theta$ to be $R_{\rm nrm}$.
We set $\frac{h}{nN}$ to be this lower bound so that the inequality (\ref{eq:h-in-terms-of-gamma}) holds, from which the result follows.
 \end{proof}

\section{\colorred{Appendix} : The q-ary Entropy and its Inverse}
In this section, we derive properties of the $q$-ary entropy function and its inverse.
Since $H_q$ is a strictly increasing concave function on $(0,\frac{q-1}{q})$, $H_{q}^{-1}$ is a strictly increasing convex function on the open interval $(0,1)$. 
Observe that for $x \in (0,1)$,
\begin{align}
H_q'(x) \hilm{\coloneqq \frac{d}{dx} H_q(x)} &=  \log_q(q-1) - \log_q x + \log_q (1-x) \label{eq:Hq'}, \\
(1-x)H_q'(1-x) 
&= H_q(1-x) + \log_q x. \label{eq:Hq2'}
\end{align}
Since $H_q(y)$ is a continuously differentiable function for $y \in (0,1-\frac{1}{q})$, by the inverse function theorem, we have that
\begin{align}
(H_q^{-1})'(y) = \frac{1}{H_q'(H_q^{-1}(y))} \label{eq:hqinvdiff}
\end{align}
for $y \in (0, 1)$,  where $(H_q^{-1})'(y) \hilm{\coloneqq \frac{d}{dy} H_q^{-1}(y)}$.
These technical properties of the $q$-ary entropy function are used to obtain Lemma~\ref{lem:nonincreasing} which pertains to the monotonicity of $\frac{1}{1-\theta} H_q^{-1} \left(\frac{1+r}{2} - r{\theta} \right)$ with respect to $\theta$, 
and Lemma~\ref{lem:continuity} which is about continuity.

Now define $f \coloneqq 1-  H_q^{-1}(\frac{1+r}{2} -r \theta)$. Observe that 
$\frac{df}{d\theta} 
= r (H_q^{-1})'(\frac{1+r}{2} -r \theta) 
= \frac{r}{H_q'(H_q^{-1}(\frac{1+r}{2} -r \theta ))}
= \frac{r}{H_q'(1-f)}$. We now introduce Lemma \ref{lem:nonincreasing} which makes an assertion on the monotonicity of the function $\frac{1-f}{1-\theta}$.
\begin{lemma}\label{lem:nonincreasing}[Monotonicity]
Let $p$ be prime, $q=  p^2$, and $r, R \in [0,1]$ such that (\ref{ineq:feasible-condition}) holds. Then 
$
\frac{1-f}{1-\theta} 
$
is a non-increasing function with respect to ${\theta \in [0,\frac{1+R}{2}]}$.
\end{lemma}
\begin{proof}
Now 
$\frac{d}{d\theta} \frac{1-f}{1-\theta}
 = \frac{1-f}{(1-\theta)^2} 
 - \frac{1}{1-\theta} \frac{df}{d\theta} $
and
 $\frac{df}{d\theta} = \frac{r}{H_q' (1-f)}$. 
 Hence 
 $\frac{d}{d\theta} \frac{1-f}{1-\theta} \le 0$ if and only if 
$(1-f)H_q' (1-f)\le r (1-\theta )$. 
From (\ref{eq:Hq2'}), we get
\begin{align}
(1-f)H_q' (1-f)
    &=  H_q(1-f) + \log_q f =  \left( \frac{1+r}{2} -r \theta \right) + \log_q f. \notag
\end{align}
Thus $(1-f)H_q' (1-f)\le r (1-\theta )$ holds if and only if 
$r \ge  1 + 2\log_q f$, the latter inequality of which holds because of (\ref{ineq:feasible-condition}).
\end{proof}
\begin{lemma} \label{lem:continuity}[Continuity]
Let $x,y \in (0,\frac{q-1}{q})$ where the integer $q$ is greater than 2 and $x>y$. 
Then
$
H_q^{-1}(y) \ge H_q^{-1}(y) - (x-y) c(q,x),
$
where our continuity constant is
\begin{align}
c(q,x)\coloneqq \left( \log_q (q-1) 
+ \log_q\left( 
        \frac{1}{ \hilm{H_{q}^{-1}}( x )}-1 \right) 
   \right)^{-1} \label{eq:continuity-constant}.
\end{align}
\end{lemma}
\begin{proof}
The convexity and continuous differentiability of $H_q^{-1}$ on the unit open interval imply that 
$  H_q^{-1}(y)  \ge  H_q^{-1}(x) - (x-y) (H_q^{-1})'(x).$
Use of (\ref{eq:Hq'}) with (\ref{eq:hqinvdiff}) then gives the result.
\end{proof}

\bibliography{mybib}{}
\addcontentsline{toc}{paper}{Bibliography}
\bibliographystyle{ieeetr}

\end{document}